\documentclass[aoas,nameyear,dvips]{arximspdf}
\usepackage{graphics}

\doi{10.1214/09-AOAS274}
\volume{3}
\issue{4}
\pubyear{2009}
\firstpage{1370}
\lastpage{1381}

\begin{document}
\begin{frontmatter}

\title{Assessing uncertainty in the American\\ Indian Trust Fund\thanksref{T1}}
\runtitle{Assessing uncertainty in the American Indian Trust Fund}
\thankstext{T1}{Work performed under Department of the Interior,
Minerals Management Service contract number GS-10F-0033M, 62817.}

\begin{aug}
\author[A]{\fnms{Edward} \snm{Mulrow}\corref{}\ead[label=e1]{mulrow-edward@norc.org}},
\author[A]{\fnms{Hee-Choon} \snm{Shin}\ead[label=e2]{shin-hee-choon@norc.org}}\and
\author[A]{\fnms{Fritz} \snm{Scheuren}\ead[label=e3]{scheuren@aol.com}}
\runauthor{E. Mulrow, H.-C. Shin and F. Scheuren}
\affiliation{National Opinion Research Center at the University of Chicago}
\address[A]{E. Mulrow\\
H.-C. Shin\\
F. Scheuren\\
National Opinion Research Center\\
4350 East West Hwy\\
Bethesda, Maryland 20814\\USA\\
\printead{e1}\\
\phantom{E-mail: }\printead*{e2}\\
\phantom{E-mail: }\printead*{e3}} 

\end{aug}

\received{\smonth{5} \syear{2009}}
\revised{\smonth{7} \syear{2009}}

%
\begin{abstract}
Fiscal year-end balances of the Individual Indian Money System (a part
of the Indian Trust)
were constructed from data related to money collected in the system and
disbursed by the system
from 1887 to 2007. The data set of fiscal year accounting information
had a high proportion
of missing values, and much of the available data did not satisfy basic
accounting relationships.
Instead of just calculating a single estimate and arguing to the Court
that the assumptions
needed for the computation were reasonable, a distribution of
calculated balances was developed
using multiple imputation and time series models. These provided
information to assess the
uncertainty of the estimate due to missing and questionable data.
\end{abstract}

%

\end{frontmatter}

\setcounter{footnote}{1}
\section{Introduction}\label{intro.ed.etal}
Starting in the later part of the 19th century, the
U.S. Department of the Interior has administered accounts of funds held
in trust for Indian tribes within Tribal Trust accounts, and for
individual Indians
within Individual Indian Money (IIM) accounts. The funds in the accounts
derive from diverse sources such as funds from litigation judgments or
settlements and funds derived from revenue producing activity on lands. There
have been numerous criticisms of the \mbox{Interior's} management of the trust fund
system over the years. In 1994 Congress enacted the American Indian Trust
Fund Management Reform Act (Pub. L. No.~103--412, 108 Stat.
4239) requiring the Interior to account for the balances of funds in
these accounts.

In June of 1996, a class action lawsuit was filed in the U.S. District
Court for the District of Columbia seeking to compel a historical
accounting of Individual Indian Money (IIM) accounts. The case is
complex, and has been in litigation for over 13 years. We will not
attempt to summarize all the events that have occurred, but Court
filings and hearing transcripts can be found at the Department of
Justice website dedicated to the\break case,
\url{http://www.usdoj.gov/civil/cases/cobell/index.htm}, as well\break as the
Plaintiffs' website, \url{http://www.indiantrust.com/}.
Additionally,\break
\href{http://indianz.com}{http://indianz.com} has many news items on the case.

The event associated with the case that is relevant to the statistical
problem that is the focus of this paper is the outcome of an October
2007 trial held to evaluate the Interior's progress toward completing
its historical accounting for IIM accounts. In its January 2008
findings of the October trial, the Court held that a historical
accounting of IIM accounts was impossible given the level of
Congressional funding, and concluded ``$\ldots$that a remedy must be
found for the Department [of the Interior's] unrepaired, and
irreparable, breach of its fiduciary duty over the last century.'' In
subsequent hearings, the Court described the remedy as determining
``$\ldots$monies that were in fact collected and made it into Treasury---into trust funds in some way, but have not been adequately accounted
for'' (March 5, 2008 Transcript of Status Conference before the
Honorable James Robertson United States District Judge).

As the statistical contractor for the Department of the Interior Office
of Historical Trust Accounting (OHTA), our approach to the problem was
to try to limit modeling assumptions and let the available data speak
for themselves. Instead of just calculating a single estimate and
arguing to the Court that the assumptions needed for the computation
were reasonable, a distribution of calculated balances was developed to
assess the uncertainty of the estimate due to missing and questionable data.

\section{Understanding the data}\label{understanding.data}
\begin{figure}

\includegraphics{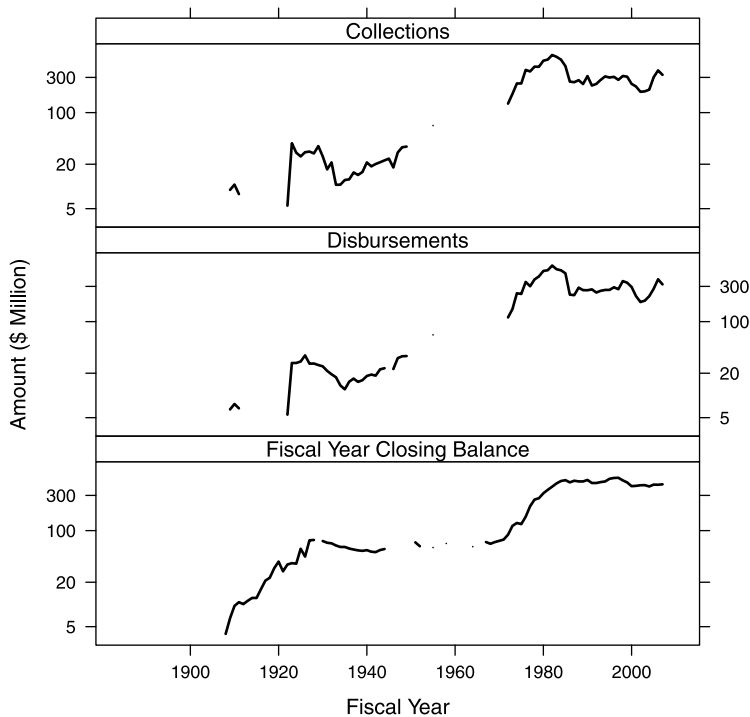}

\caption{Key accounting variables over time. Dollar amount values are
shown on a log scale.}\label{key.account.var}
\end{figure}

An Excel spreadsheet of the data that were used in our analysis of the
aggregate IIM System balance is available for download at
\url{http://www.norc.org/iim}. These include annual IIM System collections,
disbursements and balance data obtained by OHTA contractors from IIM
System government reports, and Osage headright\footnote{A ``headright''
is the right to receive a quarterly distribution of funds derived from
the Osage Mineral Estate, which is the oil, gas and other mineral
subsurface of the approximately 1.47 million acre Osage Reservation.}
data obtained from the Osage Nation website. The historical IIM System
accounting data provide a basis for analyzing IIM System information to
see if there were monies that were ``not adequately accounted for.''
Figure~\ref{key.account.var} is a graphical display of key system
accounting variables---collections, disbursements and balance data---over
the time period of interest (1887--2007). It is evident visually that a
large proportion of data are missing, and there appears also to be some
questionable observations.

The Court's view is that the issue ``$\ldots$is about dollars into the
IIM, dollars in and dollars out'' (April 28, 2008 Transcript of Status
Conference Before the Honorable James Robertson, United States District
Judge, 115 at 18). Thus, collections--dollars in and
disbursements--dollars out are two key variables, but they have a
large amount of missing information.

\begin{figure}

\includegraphics{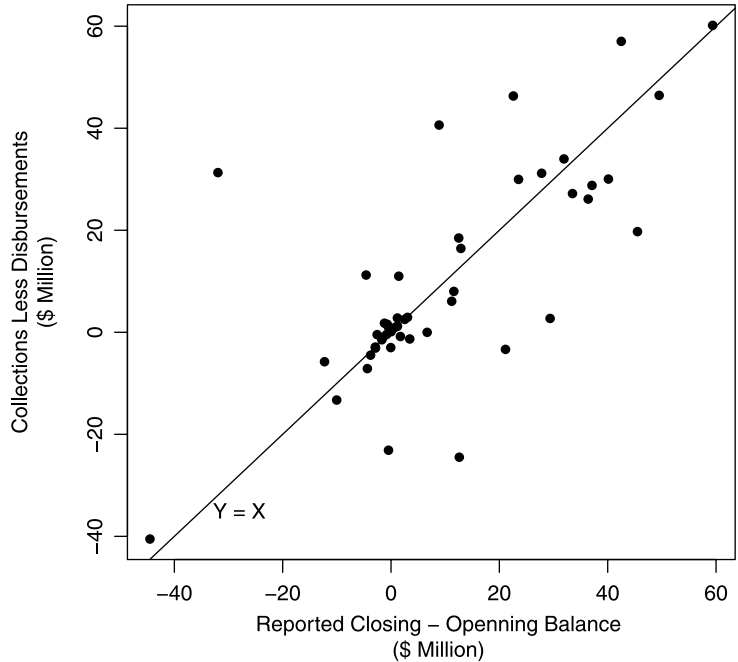}

\caption{Collections Less Disbursements vs. Change in Balance. Only
fiscal years that have observed values for collections, disbursements,
opening balance (prior fiscal year closing balance) and closing balance
are displayed. Based on accounting relationships, the data should all
fall on the $Y=X$ line. In accounting terms, the data do not ``foot.''}
\label{coll.less.disburse.vs.change.bal}
\end{figure}

A close look at the data also reveals that for years after 1911 and
before 1996, where collections, disbursements and balance data are
available, values do not ``foot,'' that is, the opening balance (prior
fiscal year ending balance) plus collections less disbursements does
not equal the ending balance. Figure~\ref{coll.less.disburse.vs.change.bal} is a graphical depiction of this:
the difference between yearly collections less disbursements is plotted
against the yearly change in balance (closing less opening balance). If
the data do foot, all the plotted points would lie on the $Y=X$ line.
But this is generally not true for these data. There may be legitimate
reasons why the numbers do not foot, for example, collections and
disbursements values come from different types of government reports
than do balance values; government reports are created for different
purposes and perhaps at different time periods. Even so, Figure~\ref{coll.less.disburse.vs.change.bal} lends credence to the Court's notion
that there may be uncertainty in the accounting of IIM System funds.

While these data have their weaknesses, they are the only IIM System
data available for determining an estimate of how much of the System
funds may be ``not adequately accounted for.'' We decided that a
two-step approach was needed to fully assess the uncertainties
exhibited in these data: (1) an imputation modeling step to assess the
uncertainty due to missing data, and (2) a synthetic modeling step to
assess the uncertainty due to accounting irregularities, which we refer
to as government reporting uncertainties.

\section{Multiple imputation modeling}\label{multiple.imputation.modeling}
After considering imputation modeling alternatives, we decided that
multiple imputation [Rubin (\citeyear{rubin87})] was applicable for this accounting
application even if its reported weaknesses [Binder (\citeyear{binder96}), Fay (\citeyear{fay96})]
were such that the methodology would lead to an overstatement of the
uncertainty; an outcome favorable to the Plaintiffs. The government,
when informed that there was such a risk, accepted our approach as the
best of available options.

Multiple imputation replaces each missing value with a set of plausible
values. The distribution of these ``plausibles'' gives us a way to
represent the uncertainty about the right value to impute. Each
completed multiply imputed data set can be analyzed using standard
procedures for complete data, and the results across these analyses
combined, so that all the uncertainty components in the analysis---model
uncertainty and missing value uncertainty---are accounted for in the analysis.

General advice is to include (within reason) as many variables as you
can in the multiple imputation model. This includes variables that are
potentially related to both the imputed variables and the missingness
of the imputed variables [Schafer (\citeyear{schafer97})]. The Osage headright
variable is related to the economic conditions that existed over time
in Indian Country, and it is known for the whole time period of
interest, so we included it in our analysis. In terms of variables that
are potentially related to the missingness of the imputed variables, we
conducted diagnostic modeling of the probability of a missing
collections or disbursements value. The diagnostics indicate that
fiscal year, Osage headright and balance are all potentially related to
the missingness of collections and disbursements [Pramanik (\citeyear{pramanik08})].

A number of other variables, for example, the portion of the balance
invested in government securities, were also available for a small
portion of the time period of interest. These data were considered,
but, ultimately, not included in our analysis. Having too many
variables with a considerable number of values missing---in fact, more
missing values than for the two variables of primary interest---might
have made it harder to construct a credible imputation model. In the
end, therefore, we proceeded with developing an imputation model based
only on fiscal year, Osage headright, balance, collections and disbursements.

We assumed that the data are from a continuous multivariate
distribution and contain missing values that can occur for any of the
variables. Furthermore, the missing values are assumed to be missing at
random [MAR, Rubin (\citeyear{rubin76})], so that the probability that an
observation is missing can depend on observed values ($\mathbf{Y}_{\mathrm{obs}}$), but not on missing values ($\mathbf{Y}_{\mathrm{mis}}$). The
effect of assuming that the missingness is entirely MAR, as is typical
in most settings [Scheuren (\citeyear{scheuren05a})], is to introduce some uncertainty
in the measurement of the uncertainty. For missing completely at random
(MCAR), using an MAR model would tend to lead to some overstatement of
the uncertainty but probably not much, assuming the variables chosen to
do the imputations are related to the variables that are missing. For
nonignorable, not missing at random (NMAR) missingness we cannot
speculate, in general, about the nature and size of any effects that
may arise. All three types are conceptually possible in any given
setting, that is, they can all be present [e.g., Scheuren (\citeyear{scheuren05b})].
However, given our belief that NMAR missingness is minimal likely for
our set of historical data, the impacts cannot be large.

A good robust multivariate model for use with multiple imputation is
the multivariate normal model with a noninformative prior [Schafer (\citeyear{schafer97})]. The complete-data posteriors, which are used to generate
imputations, are
\begin{eqnarray}
\bolds{\Sigma}|\mathbf{Y}&\sim& W^{-1}\bigl(n-1,(n-1)\mathbf{S} \bigr)\nonumber,\\
\bolds{\mu}|\bolds{\Sigma},\mathbf{Y}&\sim& N\biggl(\bar{\mathbf{y}},\frac{1}{n}\bolds{\Sigma} \biggr)\nonumber,\\
\mathbf{Y}_i'|\bolds{\mu},\bolds{\Sigma}&\sim& N(\bolds{\mu}
,\bolds{\Sigma})\qquad\forall i=1,\ldots,n,\nonumber
\end{eqnarray}
where $W^{-1}$ denotes the inverse Wishart distribution, $\mathbf{Y}$ is the completed data matrix (which is composed of the observed
values, $\mathbf{Y}_{\mathrm{obs}}$, and the filled-in missing values,
$\mathbf{Y}_{\mathrm{mis}}$), $\mathbf{Y}_i$ is a row in the data matrix,
$n$ is the number of years in the data matrix to be completed,\footnote{The starting data matrix has 128 rows; one for each year in the
1880--2007 timeframe, but the focus of our analysis is for the
1887--2007 timeframe, which is 121 years. Because of the time series
modeling described in Section~\ref{synth.modeling.time} of this paper,
we needed to go back to 1880 in order to forecast values starting in
1887.} $\mathbf{S}$ is the sample covariance matrix, $\bar{\mathbf{y}}$ is the sample mean vector, and $N(\cdot)$ denotes the
multivariate normal distribution.

The imputation of the missing annual collections, disbursements and
balance values for the IIM System was completed using the SAS MI
procedure, which uses an MCMC algorithm to generate observations from
the posterior distribution. The SAS program that implements this can be
found at \url{http://www.norc.org/iim}. The imputation of each missing
collections, disbursements and balance value was done 10,000 times.

The multiple imputation literature, written 30 years ago during an era
of expensive computing, generally suggests that 3-to-5 imputations
would be sufficient for assessing the contribution to an estimated
value's uncertainty due to missing information. The theory behind this
relies on the use of a multivariate normal distribution, which is also
the basis for our imputation modeling. Since we live in an era of less
expensive computing, we chose to use a much larger number of
imputations, 10,000. Our computing power was sufficient for this many
imputations, and our data matrix was not so large that it would have
taken an inordinate amount of time to complete the imputation process.

\section{Synthetic modeling for addressing the uncertainty inherent in available data}\label{synth.modeling.time}
The footing errors for the available data from 1908--1995 (see
Figure~\ref{coll.less.disburse.vs.change.bal}), and under-reporting
issues for the 1922--1949 collections and disbursements data,\footnote
{Collections and disbursements values from 1922--1949 came from
``Statement of Money Received and Expended by Disbursing Agents of the
Indian Service Without being Paid into General Treasury of the United
States'' reports. Thus, we know that monies held within the Treasury
were not accounted for in these reports.} lead us to conclude that the
pre-1996 data are questionable. Therefore, we feel that the results of
our analysis should reflect more uncertainty than if all the data used
for modeling were thought to be reliable.

To introduce this uncertainty into the data so that it would be
reflected in the confidence bounds, we first fit a model to all the
annual data (1880--2007) for each of the 10,000 imputed data sets. We
then created a realization from each model for the years in question
(1887--1995), which included a random noise component. This provided us
with 10,000 ``synthetic'' data sets. We use the term ``synthetic'' here
because the processing steps we have used are similar to the creation
of synthetic data, as described by Reiter (\citeyear{reiter02}). As noted by Rubin (\citeyear{rubin93}), the result of using this type of modeling approach will
still produce valid statistical inference, but the variance will be
larger than the variance from the original data ``$\ldots$because there
is a reduction in information relative to the actual microdata$\ldots.$''

Given that we have annual accounting observations in each of our 10,000
complete data sets, a natural way to model the data is through time
series techniques. At this point it is important to recall that our
goal is to estimate the 2007 year-end balance by estimating total
collections and total disbursements over the time period 1887--2007,
and then taking the difference between the total collections and total
disbursements. Therefore, we restricted out attention to just the
collections and disbursements variables.

In using time series techniques for modeling these data, we must not
only take into account the serial correlation within each of the
collections and disbursements series, we must also take into account
the cross-correlation between the two variables---both contemporaneous
and prior value correlations. Vector autoregressive moving average
(VARMA) processes are a class of models that handle such correlation
structures. Following Brockwell and Davis (\citeyear{brockwelldavis87}), $\{\mathbf{X}_t,
t=0,\pm1,\ldots,\}$ is a bivariate VARMA($p,q$) process if $\{
\mathbf{X}_t\}$ is stationary and
\begin{eqnarray*}
\mathbf{X}_t=\mathbf{M}+\sum_{i=1}^{p}\bolds{\Phi}
_{i}\mathbf{X}_{t-i}+\mathbf{Z}_t+\sum_{i=1}^{q}\bolds{\Theta}
_{i}\mathbf{Z}_{t-i},
\end{eqnarray*}
where $\mathbf{X}_t = (X_{t1}, X_{t2})'$ and $\mathbf{Z}_t =
(Z_{t1}, Z_{t2})'$ are series of bivariate vectors, $\mathbf{M} =
(M_1, M_2)'$ is a bivariate constant (mean) vector, $\bolds{\Phi}
_1,\ldots,\bolds{\Phi}_p$ and $\bolds{\Theta}_1,\ldots,\bolds{\Theta}_q$ are $2\times2$ matrices, and $\mathbf{Z}_t\sim
\mathit{WN}(\mathbf{0},\bolds{\Sigma})$, a bivariate white-noise process
with common $2\times2$ covariance matrix $\bolds{\Sigma}$.

\begin{figure}[b]

\includegraphics{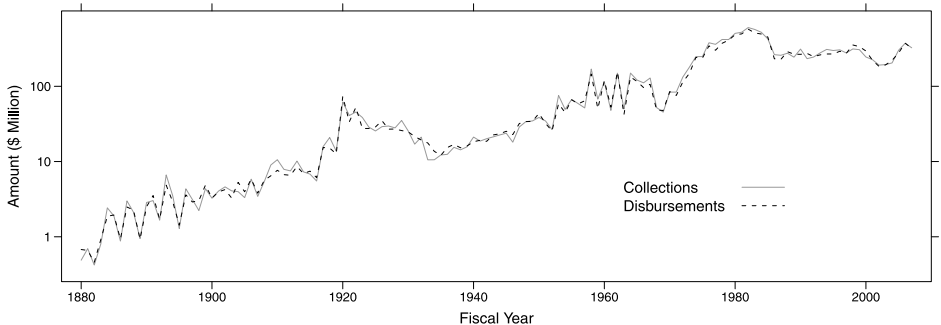}

\caption{Example collections and disbursements time series after
imputation of missing values.
Dollar amount values are shown on a log scale. The plot's aspect ratio
has been chosen using the
45 banking rule developed by Cleveland (\protect\citeyear{cleveland93}).}
\label{log.coll.dis.timeseries}
\end{figure}

A basic assumption for this type of model is that the time series
process is stationary. Figure~\ref{log.coll.dis.timeseries} shows time
series plots the log transformed collections and disbursements data for
one of the 10,000 bivariate time series generated from the multiple
imputation process. It is plausible that log collections and log
disbursements time series are stationary, or stationary after removing
an increasing trend over time, and we proceeded to fit VARMA models to
the transformed series, that is, $X_{t1} = \operatorname{Log}$(Collections) and
$X_{t2} = \operatorname{Log}$(Disbursements) for years $t = 1880,\ldots,2007$.

For each of the 10,000 bivariate time series, we need to estimate the
unknown coefficients $\mathbf{M}$, $\bolds{\Phi}_i$ and
$\bolds{\Theta}_i$, and then use the fitted model to create a
different realization from the time series model. It would have been
overly time consuming to check for stationarity and fit the ``best''
VARMA model to each of the 10,000 time series produced from the
multiple imputation procedure. So we checked a small set time series
from the 10,000, and ran some high level diagnostics on a larger subset.

We checked the small subset of times series using the tentative order
identification routines found in
Spliid (\citeyear{spliid83}),
Koreisha and Pukkila (\citeyear{koreishapukkila89}), Quinn (\citeyear{quinn80}),
which are based on identifying the $p$ and $q$ orders that
minimized a statistical information criterion. A VAR process with order
$p$ between 2 and 5 was consistently identified as the VARMA process
that produced the minimum AICC value. In order to produce realizations
of a VAR($p$) model for the time period of interest (1887--1995), we
needed to have the starting time series go back further in time (before
1887) by $p$ years. Of the data available to us before we started the
multiple imputation process, the Osage data went back furthest in time
to 1880. So, we had 7 years of available data that predated the time
period of interest. Therefore, the highest order value we could choose
that would not predate our available data was $p$ = 7. We settled on
fitting a VAR(7) process to each of the 10,000 series. We used the VAR
models fit to each series to generate synthetic collections and
disbursements values for each fiscal year in the 1887--1995 time
period. The SAS program found at \url{http://www.norc.org/iim} provides
the details of how this was implemented.

\begin{figure}

\includegraphics{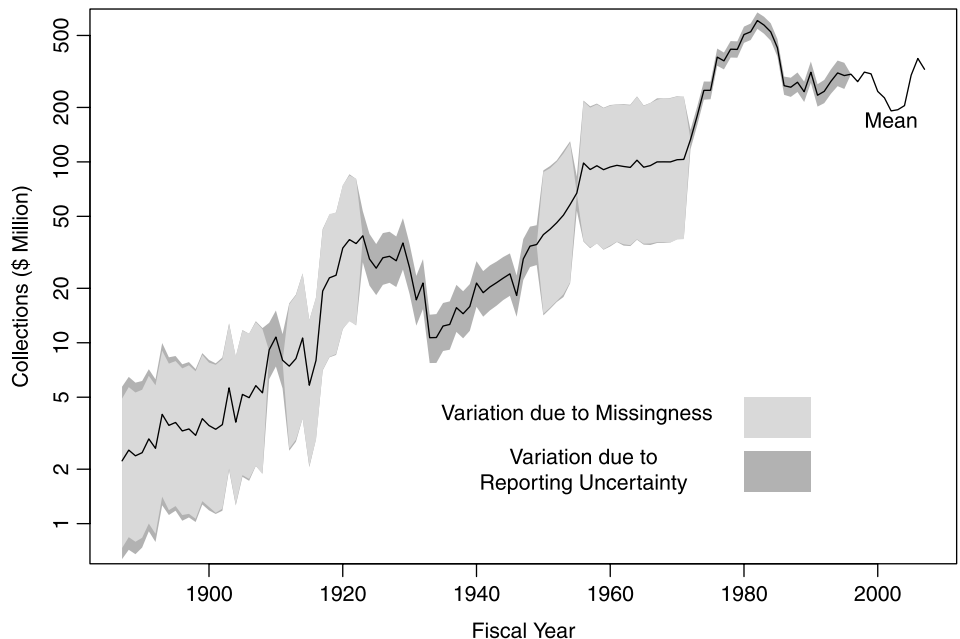}

\caption{Collections, on a log scale, variation due to missingness
and variation due to reporting uncertainty over time. The trend line
and the shaded region represent the mean and a 95 percent confidence
interval, respectively, for each fiscal year. Variation due to
missingness comes from the multiple imputation model. Variation due to
reporting uncertainty is incorporated by a synthetic model based on the
multiple imputation results.}\label{coll.95pct.ci}
\end{figure}

Figure~\ref{coll.95pct.ci} shows the mean value of collections for each
fiscal year of the 10,000 values assigned to the year from the modeling
process, and 95 percent confidence intervals are shaded in two ways to
provide a visual of the uncertainty of the collections values. The
lighter shading shows the variation of the imputed values from the
initial multiple imputation modeling. The darker shading shows the
additional variation added to the collections values due to the
reporting uncertainty. We see the largest amount of variation present
in the time periods where collections values were missing. For time
periods with questionable reported data, a relatively small amount of
variation has been added for fiscal years immediately preceding 1996,
but the additional variation due to reporting uncertainty gets larger
as we go further back in time. Similar features are found in an
analogous graph of disbursements (not shown).

For each completed data matrix, total collections and total
disbursements across the 121-year period between 1887 and 2007 were
calculated. The difference between these two values, which we will
refer to as the ``calculated balance,'' was found for each of the 10,000
completed data matrices. The calculated balance distribution is fairly
symmetric about the median value of \$580.4. The 25th and 75th
percentiles of the distribution are \$502.0 million and \$661.7
million, respectively.

\section{Discussion}\label{discussion.ed.etal}
We developed a methodology for estimating the 2007 fiscal year-end
balance of the IIM System based on estimating the total amount of money
collected in the system and disbursed by the system from 1887 to 2007.
Because the available annual collections and disbursements data were
not available for about one-third of the years in the time frame, we
used a multiple imputation methodology to fill-in the missing values.
Additionally, many of the reported collections and disbursements values
exhibited questionable behavior, and in some cases were known to be
underreported. Therefore, our approach for determining the balance for
the 1887--2007 time frame concentrated on assessing the distribution of
possible balance values, which provides an evaluation of the
uncertainty of the estimated balance due to missing information and
reporting uncertainty in the available data.

Our final assessment is that the calculated balance for the 1887--2007
time frame has an average value of \$583.6 million, which is \$159.9
million higher than the stated 2007 balance of \$423.7 million.
However, the distribution of the calculated balances has large
variation, as exhibited by a 95 percent confidence interval that ranges
from a lower bound of \$353.1 million to an upper bound of \$833.5 million.

The calculated balance distribution does not reflect any inflation or
interest adjustment on the dollar amounts. The Court ruled that it
could not award interest in the U.S. District Court. Generally, only
the Court of Federal Claims may award interest in a suit against the
Government.\footnote{This is a legal issue that is not simple to
explain. We refer the interested reader to the August 7 2008, United
States District Court for The District Of Columbia, Cobell v.
Kempthorne Memorandum, which can be found at 
\href{http://www.usdoj.gov/civil/cases/cobell/index.htm}{http://www.usdoj.gov/civil/cases/}
\href{http://www.usdoj.gov/civil/cases/cobell/index.htm}{cobell/index.htm}.}

The uncertainty reflected in the distribution of the calculated balance
is a result of the government's inability to find a consistent set of
documents that shows the IIM system balances over the time frame of
interest. In other such circumstances, the government compensates those
on whom uncertainty is imposed by choosing a point on the distribution
favorable to the other party. For example, the Internal Revenue
Service\footnote{IRS Internal Revenue Manual, 4.47.3.3.1.}
uses the 95th percentile of the distribution, which means that the
taxpayer or person being audited is 95\% sure of not overpaying.
DHHS\footnote{DHHS Centers for Medicare and Medicaid Services, Program
Integrity Manual, Section~3.10.5.1.} uses the 90th percentile in
similar circumstances, which is slightly less favorable to those being audited.

In its August 8, 2008 memorandum following the June 2008 Cobell v.
Kempthorne trial, the Court found that our model (the government's
model) was imperfect, but that it presented ``$\ldots$a plausible
estimate of funds withheld,'' and that it was ``$\ldots$useful in
evaluating the uncertainty in the existing trust data,'' particularly
the ``overall uncertainty at the balance level.''

The Court chose to use the 99 percentile of the calculated balance
distribution, \$879.3 million, as a point on the distribution favorable
to the plaintiffs. This more conservative limit was chosen because
``$\ldots$there is more uncertainty in the data$\ldots$ historical
reports are not biased but may be understated, [Integrated Records
Management System] data has important reliability problems, and the
qualified audit data is, after all, only qualified, and was not even
subjected to the time-series remodeling step.'' Adjusting for the stated
fiscal year-end IIM system balance of \$423.7 million produces a \$
455.6 million understatement of the system balance, which was awarded
to the Plaintiffs.\footnote{Plaintiffs have appealed this decision. In
particular, the Plaintiffs argue that the judgment should include
interest because the government benefited over the years from having
extra money in the US Treasury. If the Appeals Court rules in the
Plaintiffs' favor on the interest issue, an interest-adjusted,
calculated balance distribution can be derived by applying agreed-upon
annual interest rates to each of the 10,000 time series.}

\section{Alternate modeling approaches}
The timeline we were given to complete the analysis of the IIM System
data was short. While we believe that the uncertainty modeling
presented to the court was appropriate, hindsight suggests a number of
competing models. For example, we did not include a trend term in the
time series model. Even though the diagnostic checks we performed
suggested that a VAR(7) model was a reasonable choice, Figure~\ref{log.coll.dis.timeseries} suggests an increasing trend over time. We
have rerun the model with a linear time trend, and have found that the
distribution of calculated balances does not change appreciably. But,
another alternative that we have not investigated is to use
differencing of the collections and disbursement data to remove trends.

Given that the multiple imputation modeling is a Bayesian hierarchical
model, maybe we should have used a Bayesian vector autoregressive model
[BVAR Litterman (\citeyear{litterman86}), Brandt and Freeman (\citeyear{brandtfreeman06})] to incorporate the
additional uncertainty in the calculated balance distribution. But it
is not clear to us that this was needed. The validity of multiple
imputation does not require one to fully subscribe to the Bayesian
paradigm [Rubin (\citeyear{rubin87})], so it is not clear that we needed to use a
Bayesian model for the second stage. We have also questioned whether
the two-step model should have been completed in one modeling process.
A BVAR could be included in the multiple imputation hierarchical model,
possibly along with a measurement error model [Ghosh, Sinha and Kim (\citeyear{ghoshetal06})] for
the collections and disbursement observation. We have not yet attempted this.

We encourage others to explore these data and suggest ways in which
they can be analyzed. The data set, as noted, is available for download
at \url{http://www.norc.org/iim}. We would be interested to know if
other modelers using all of the available data, plus perhaps additional
economic indicators, provide estimates that are consistent with our own.

\begin{supplement}
\sname{Supplement A}\label{supp.data.ed}
\stitle{Data Set}
\slink[doi]{10.1214/09-AOAS274SUPPA}
\slink[url]{http://lib.stat.cmu.edu/aoas/274/Supp\%20A\%20-\%20IIM\%20System\%20Uncertainty\%20Modeling\%20Data.xls}
\sdescription{The data set (IIMSystemUncertaintyModelingData.xls) is
available for download at \url{http://www.norc.org/iim}.}
\sdatatype{.xls}
\end{supplement}
\begin{supplement}
\sname{Supplement B}\label{supp.code.ed}
\stitle{SAS Code}
\slink[doi]{10.1214/09-AOAS274SUPPB}
\slink[url]{http://lib.stat.cmu.edu/aoas/274/Supp\%20B\%20-\%20IIM\%20System\%20Uncertainty\%20model.sas}
\sdescription{The SAS program that we used to read the input data,
apply the modeling methodologies, and produce the outputs used in
summaries and graphs is available at \url{http://www.norc.org/iim}.}
\sdatatype{.xls}
\end{supplement}

\printaddresses

\end{document}